# Information systems security in special and public libraries: an assessment of status

**Roesnita Ismail[1] and A.N. Zainab[2]**
[1]Faculty of Science & Technology, Islamic Science University Malaysia,
Bandar Baru Nilai, 71800 Nilai, Negeri Sembilan, MALAYSIA
[2]Digital Library Research Group,
Faculty of Computer Science & Information Technology,
University of Malaya, 50603 Kuala Lumpur, MALAYSIA
e-mail: roesnita@yahoo.com; zainab@um.edu.my

## ABSTRACT

*Explores the use of an assessment instrument based on a model named library information systems security assessment model (LISSAM) to assess the ISS status in special and public libraries in Malaysia. The study aims to determine the implementation status of technological and organizational components of the LISSAM model. An implementation index as well as a scoring tool is presented to assess the IS safeguarding measures in a library. Data used was based on questionnaires distributed to a total of 50 individuals who are responsible for the information systems (IS) or IT in the special and public libraries in Malaysia. Findings revealed that over 95% of libraries have high level of technological implementation but 54% were fair poorly on organizational measures, especially on lack of security procedures, administrative tools and awareness creation activities.*

## INTRODUCTION

Information system security (ISS) practices encompass both technical and non-technical issues to safeguard organizational assets from a variety of threats. Information System (IS) in libraries support the delivery of image, services and collections to local and remote patrons and this availability over the Internet inevitably exposes it to security threats. As a result some controls need to be in place to secure the IS (Guttman and Roback 1995; Westby and Allen 2007; Gupta and Sharman 2008; Scarfone et al. 2008).

Studies on ISS practices in libraries are few and therefore one could not gauge whether the library sector is lacking or adequate in IS security management (Newby 2002). This lack of coverage is surprising since information is a library's core business and information delivered over an information system requires consistent monitoring and maintenance in order to avoid possible intrusion. This study attempts to propose an instrument to assess the level of implementations of ISS status in Malaysian libraries.

## RELATED STUDIES IN MALAYSIA

The few Malaysian-related studies covered mainly ISS in health care, IT organizations, and government sectors. Al-Salihy, Ann and Sures (2002) assessed the effectiveness of ISS in a Malaysian IT organization. Their findings revealed that disincentives certainty, systems environment and security, software control, and organizational maturity are key factors contributing to ISS' effectiveness, while the effect of code of ethics and top management support were insignificant. Suhazimah (2007) explored ISS management in Malaysian public services (MPS) and found that awareness about information security existed in many of the MPS' organizations and the respondents believed that information security management practices were documented and had been communicated within their organizations. On the other hand, Samy, Rabiah and Zuraini (2009) studied the potential threats that exist in Malaysian health care information systems. Their study revealed that power failure was the most critical threat for the health care information systems, followed by acts of human error, technological obsolescence, hardware problems, software failures, network infrastructure failures, and malware attacks. This research holds significant value in terms of providing a complete taxonomy of threat categories in health care information systems and identifying the overall risks in the health care domain. Realizing the lack of research in the library settings, we are motivated to explore measurements that can be used to assess the ISS practices in Malaysian libraries.

## OBJECTIVES

This exploratory study aims to firstly, identify a suitable IS security model that can be adopted and adapted to formulate an assessment instrument in the library context; secondly, to determine the validity and reliability of the instrument; and finally, to assess the current IS security status in Malaysian special and public libraries. The assessment tool specifically assess the types and extent of IS safeguarding measures in a library in term of technological (security of hardware, software, workstations, networks, servers, data and its physical facilities) and organizational measures (security policy, procedures and controls, tools and methods and awareness creation activities). The geographic context of this pilot study is selected Malaysian special and public libraries.

## METHODOLOGY

This study encompassed two phases. The first phase involved identifying from literature the relevant model that could be adapted to frame the components that will be used in the survey instrument to assess the implementation status of ISS in Malaysian academic libraries (Appendix 1). The second phase involved piloting the survey instrument to determine its reliability and validity and to score the results using an instrument adapted from the *Information security Governance (ISG) Assessment Tool for Higher Education* (EDUCAUSE/Internet 2 Security Task 2004). The scores obtained would place the security measures used in libraries at very high, high, medium, low or very low levels of implementation. This will enable libraries to identify their strong and weak areas.

The survey instrument was used to collect data for analysis. The study uses a convenient sampling method to pilot the instrument on selected special and public libraries based on the minimum selection criteria, that each of these libraries must have an automated library system, provide Internet access and online services. The respondents sample consisted of 110 individuals who are responsible for the IS or IT in the special or public libraries in Malaysia. The background information on these respondents was obtained from their university library websites. Each respondent was given a self-administered questionnaire either by post, hand or e-mail attachments depending on the convenience of the locations. Each questionnaire booklet or email attachment was attached with an introductory letter which explains the research objectives, instructions and definitions of key terminologies together with self-addressed stamped envelope. Follow-up telephones calls and reminders via e-mails were made to increase the response rate. The findings were based on 50 usable questionnaire returns. Respondents came from 40 (80%) special libraries and 10 (20%) public libraries. The results of pilot survey are subjected to validity and reliability testing using the Cronbach's alpha score and subsequently measured using the ISG assessment tool to determine the implementation status of each component of the security measures as well as general performance in general.

## THE LIBRARY INFORMATION SYSTEMS SECURITY ASSESSMENT MODEL (LISSAM)

A search in literature for an ISS assessment model in libraries could not be located. As a result we looked for a suitable model in the IS domain to find one that is generic enough to be able to apply to a library situation. We found this in Hagen, Albrechtsen and Hovden's (2008) model which presented an organizational IS Staircase Model, which illustrated that an effective ISS should be built like a staircase of combined components, which are mutually dependent on each step. The two main components are the technological and organizational measures (Sundt 2006; Berghel 2005). The combined measures are formulated based on Von Solm's (2000) suggestion that the basis of an ISS should include organizational, legal, institutional aspect and the applications of best practices and security technologies. The model works on the premise that the critical foundation of an ISS is its technological infrastructure that must be in place first and foremost. The organizational components comprises four factors; the existence of an information security policy, the emplacement of procedures and controls, the formulation of adequate administrative tools and methods and awareness creation. The steps in the staircase follow a logical order so as to achieve the three primary goals of a good ISS practices, which was to ensure and protect the confidentiality, integrity and availability of an IS system (Eisenberg and Lawthers 2005). These three goals become the basis for ISS that protects the system from different security threats. We adopted and adapted the Hagen, Albrechtsen and Hovden model to derive an ISS assessment model in the library context. The library ISS refers to the means and ways a library protects the information processed by an IS and of the IS itself, which includes giving access to authorised users (confidentiality), protecting against unauthorised changes (integrity) and making the system always available and usable (availability) whenever it is needed. The proposed Library Information Systems Security Assessment Model (LISSAM) is given in Figure 1.

In the LISSAM model the higher the position on the staircase, the more complete and complex is the state of the ISS management in a library (Hagen, Albrechtsen and Hovden 2008). The first staircase illustrates that the library security environment which includes the technological foundation must always be in place. Next, the security policies must be grounded with developed rules, guidelines and plans. The third staircase illustrates that the ISS processes and procedures are supported by appropriate tools and methods. When these formal systems are implemented, the library subsequently needs to deal with the human issues who give life to the administrative security routines by applying them in their day to day activity. This final step involves raising awareness and educating everyone who is using the library IS as users and this represent the most complex and important part in the IS systems.

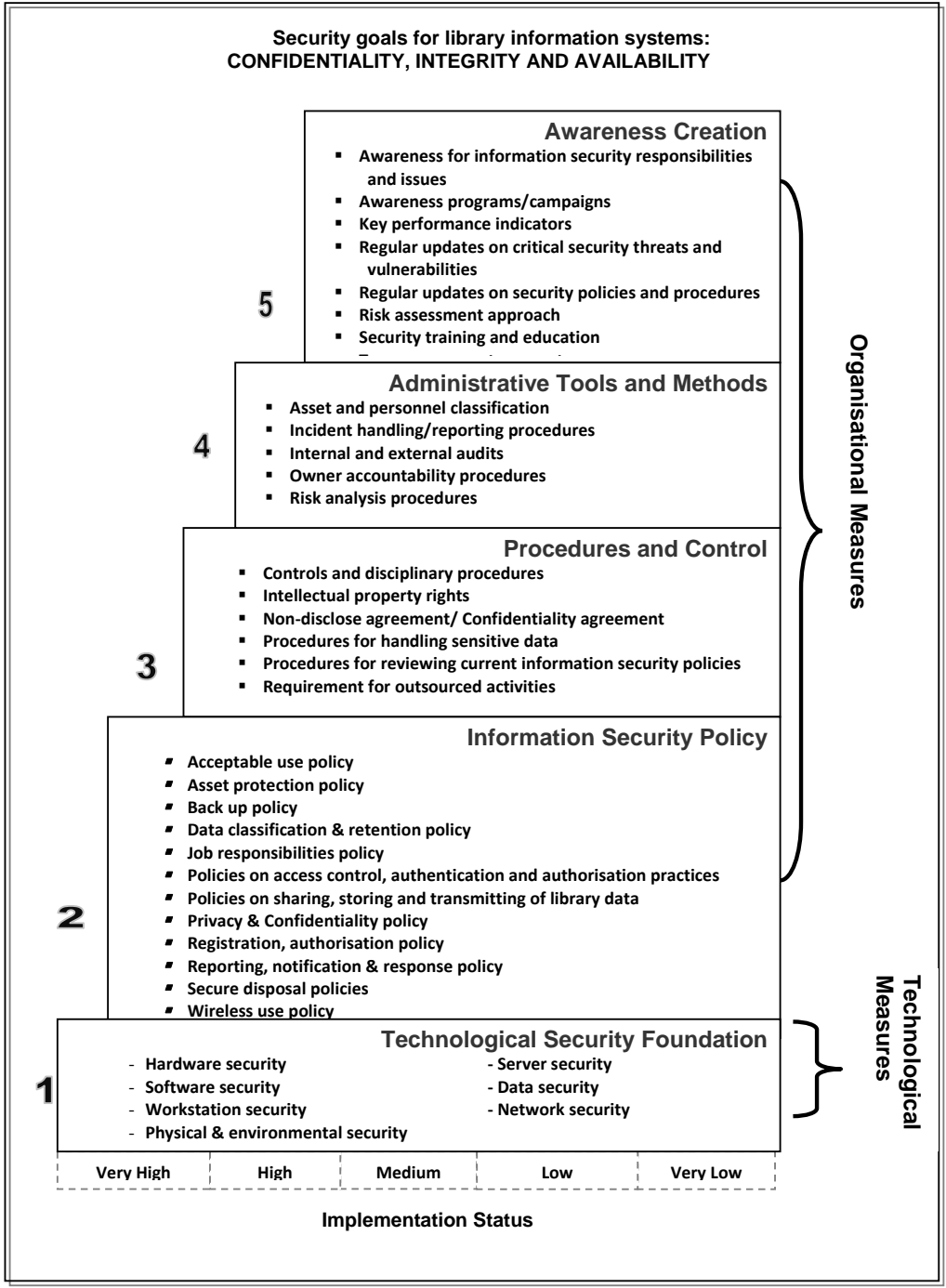

Figure 1: Library Information Security System Assessment Model (LISSAM)

The sequence of steps shown in the model illustrates the ideal sequence of combined security measures for the library IS. However, there are possibilities that some measures were more fully implemented than the others. The implementation index based on the Information Security Measure Benchmark (Information-Technology Promotion Agency, 2008) was applied to the model to create an instrument that can be used to assess the level of ISS implementation in a library. At each level, the variables are measured based on five status of implementation scores (1 = Not Implemented to 5 = Fully Implemented) that reflect the attributes of implementation (Table 1).

This study used the components factor analysis technique as suggested by Malhotra (2004) for the purpose of items reduction in the survey instrument. Results of the factor analysis (see Table 2) indicated that item commonalities scores were high as they were all above 0.60. MacCallum et. al. (1999), MacCallum et. al. (2001) and Zhao (2009) advised researchers to check commonality of each variable and drop variables that fall below 0.6. The Cronbach's alpha score is used to evaluate the consistency of the responses for each item within the instrument. The reliability scores justified the use of all the items under the 5 constructs. In this study we choose the reliability coefficient of 0.60 or higher as "acceptable".

Table 1: Levels of ISS Implementation Measures in Libraries

| Level | Status | Description of the attributes of information systems security practice |
|-------|--------|-----------------------------------------------------------------------|
| 1 | Not Implemented | No security measure has been established |
| 2 | Only some part has been implemented | Only some part of security measure has been implemented |
| 3 | Implemented but has not been reviewed | Implemented but the stage has not been reviewed |
| 4 | Implemented and reviewed on regular basis | Implemented and the state reviewed on regular basis. |
| 5 | Implemented and recognised as good example for other libraries | Implemented enough to be recognised as good example for others libraries |

(Source: Information-Technology Promotion Agency 2008)

## (a) Technological Mechanisms: Step 1

Technical security mechanisms are used to safeguard the library IS integrity, confidentiality, and availability. These include the mechanisms that are put in place to protect, control and monitor information access as well as prevent unauthorized access to data that is transmitted over a library system. The assumption is that a technological foundation must always be in place in any ISS environment and treated as the main defensive system. Moreover, the traditional ISS has emphasized on security technologies as the basis of a security system (Siponen and Oinas-Kukkonen 2007). In this study, the technological security foundation refers to the security of hardware, software, workstations, networks, servers, data and its physical facilities and environment in libraries. Technological mechanisms comprise the following issues:

i) Hardware Security

Hardware in this context refers to telephone lines, input/output ports, modems, network cablings, scanners, printers and storage media. These equipments need to be secured from any threats including thefts, power failures, equipment incompatibilities, careless damage and ensure the availability, confidentiality and integrity of data in a library (Yeh and Chang 2007; INTOSAI 1995).

ii) Software Security
Software refers to library systems, OPACs, online databases and resources made availavle over the Internet. The scope of software security encompass protecting the software components from breaches (Eisenberg and Lawthers 2005; Yeh and Chang 2007; Newby 2002).

iii) Workstation Security
The most common library's workstation security threats come from the Internet and from the users such as viruses and worms, theft and unauthorised access (Eisenberg and Lawthers 2005; INTOSAI 1995).

iv) Network Security
The network security for a library would need to disallow access to the IS from unauthorised users, while simultaneously ensuring full access to legitimate users (Eisenberg and Lawthers 2005; Yeh and Chang 2007).

v) Server Security
Libraries need to take steps to secure the email and web server applications from any intrusion, hardware or application failure due to viruses, hackers and natural disasters. The availability, confidentiality and integrity of the library server can be assured via proper implementation of specific counter measures (Eisenberg and Lawthers 2005).

vi) Data Security
The library needs a sound data management system to assure the security of its data against accidental loss, unauthorised modifications and access by taking appropriate measures (Yeh and Chang 2007; Thiagarajan 2003; Powell and Gillet 1997).

vii) Physical Facilities and Environmental Security
The term physical and environmental security refers to measures taken to protect the library systems, buildings and related supporting infrastructures or resources (including air conditioning, power supply, water supply and lighting) against physical damage associated with fire, flood and physical intrusion (INTOSAI 1995; Yeh and Chang 2007).

The Cronbach's alpha scores of all items under (i) to (vii) are above 0.60 and an overall score of 0.682 indicating high degree of internal consistency and reliability of the items listed to measure technological implementations (Table 2).

## (b) Information Security Policy: Step 2

Information security policy refers to written documentations in the form of policies and guides that laid out the process of protection linked to the overall security strategy of the library (Hone and Elloff 2002; Doherty and Fulford 2006). A security policy is needed in a library as it provides continuity, consistency and a basis for enforcing staff and patron conduct when using the library IS (Williams 2001). In order to be more practical and implementable, policies are defined by standards, guidelines and procedures (Weise and Martin, 2001). This component that received the highest Cronbach's alpha score among the 5 factors was information Security Policy which ranged from 0.849 (lowest) to 0.971 (highest) and with a total item average score of 0.844 (Table 2).

## (c) Procedures and Controls: Step 3

Procedures are step-by-step instructions on how to implement and enforce policies in the organisation (Conklin et al. 2005). Procedures and controls are implemented through work processes and procedures, which outline how resources are protected. This step consists of documents guiding individuals and organisations through user instructions, security plans, non-disclosure agreements and follow-up activities of the documented systems. The items in this factor receive Cronbach's alpha scores of between 0.70 and 0.90 and a total average of 0.671 indicating that the reliability coefficient scores is acceptable (Table 2).

### (d) Administrative Tools and Methods: Step 4

Administrative tools and methods are both proactive and reactive means in ensuring the security of IS in a library which includes asset classification, risk analysis, audits and incident reporting systems. The items in this factor receive Cronbach's alpha scores of between 0.640 and 0.956 and a total average of 0.705 indicating that the items are acceptable (Table 2).

### (e) Awareness Creation: Step 5

This step refers to the process of making people understand and aware of the importance of security, the use of security measures, the implications of security on their ability to perform their jobs and the process of reporting security violations (Pipkin 2000). The Cronbach's alpha scores for all items are above 0.60 and found to be reliable (Table 2).

Table 2: Cronbach Alpha Scores for the Five Components in the LISSA Model

| Items | Components | Cronbach's Alpha |
|-------|-----------|------------------|
| 1.0 | Technological Component | .682 |
| 2.0 | Information Security Policy | .844 |
| 3.0 | Procedures and Controls | .671 |
| 4.0 | Administrative Tools and Methods | .705 |
| 5.0 | Awareness Creation | .755 |

### THE SCORING TOOL

A scoring tool adaptation from the *Information Security Governance (ISG) Assessment Tool for Higher* Education (EDUCAUSE/Internet2 Security Task 2004) was used to measure the status of implementation for all 5 stages of LISSAM. Table 3 indicates that the total score from each of the five sections should be entered into the corresponding box ✓on this chart to determine the overall status of library's ISS practices. Firstly, the library's total score for presence of technological measures is evaluated by summing all the seven technological components (Table 3).

The library's total score for presence of organizational measures is calculated as the sum value for the presence of IS policy, procedures and controls, administrative tools and awareness creation (Table 4).

Next, the overall assessment score for the library ISS practices is evaluated based on the status of technological measures and organizational measures, which indicates implementation status of very poor practices, poor practices, average practices, good practices, or very good practices (Table 5).

Table 3: Total Score for Technological Measures

| TECHNOLOGICAL MEASURES | | Low | High | Presence |
|---|---|---|---|---|
| Total Score for Presence of Hardware Security | ✓ | 0 | 2 | Very Low |
| | | 3 | 5 | Low |
| | | 6 | 10 | Medium |
| | | 11 | 15 | High |
| | | 16 | 20 | Very High |
| Total Score for Presence of Software Security | ✓ | 0 | 10 | Very Low |
| | | 11 | 20 | Low |
| | | 21 | 40 | Medium |
| | | 41 | 60 | High |
| | | 61 | 80 | Very High |
| Total Score for Presence of Workstation Security | ✓ | 0 | 2 | Very Low |
| | | 3 | 6 | Low |
| | | 7 | 12 | Medium |
| | | 13 | 18 | High |
| | | 19 | 25 | Very High |
| Total Score for Presence of Network Security | ✓ | 0 | 5 | Very Low |
| | | 6 | 11 | Low |
| | | 12 | 22 | Medium |
| | | 23 | 33 | High |
| | | 34 | 45 | Very High |
| Total Score for Presence of Server Security | ✓ | 0 | 6 | Very Low |
| | | 7 | 12 | Low |
| | | 13 | 25 | Medium |
| | | 26 | 37 | High |
| | | 38 | 50 | Very High |
| Total Score for Presence of Data Security | ✓ | 0 | 8 | Very Low |
| | | 9 | 16 | Low |
| | | 17 | 32 | Medium |
| | | 33 | 48 | High |
| | | 49 | 65 | Very High |
| Total Score for Presence of Physical Security | ✓ | 0 | 5 | Very Low |
| | | 6 | 11 | Low |
| | | 12 | 22 | Medium |
| | | 23 | 33 | High |
| | | 34 | 45 | Very High |
| **TOTAL SCORE FOR PRESENCE OF TECHNOLOGICAL MEASURES**  (Presence of Hardware, Software, Workstation, Network, Server, Data and Physical Security) | ✓ | 0 | 41 | Very Low |
| | | 42 | 82 | Low |
| | | 83 | 165 | Medium |
| | | 166 | 247 | High |
| | | 248 | 330 | Very High |

## RESULTS OF THE PILOT SURVEY

This preliminary study reveals a general picture on the status of information systems security measures in the pilot sample of special and public libraries in Malaysia. The libraries demonstrated a higher level of implementation of technological measures than organizational measures. The results also revealed that the libraries were focusing more on the 'visible' prevention measures exemplified by observable physical aspects of monitoring tools such as security cameras, locks, warning signs and fences which are more observable by library staffs, users and outsiders as compared to the less visible security controls such as implementation of policies, procedures and awareness programmes for staffs.

### (a) Level of Implementation of Technological Measures
The level of implementation of technological measures is high for 95% in special libraries and 100% in public libraries in Malaysia (Table 6).

Table 4: Total Score for Organizational Measures

| ORGANIZATIONAL MEASURES | | Low | High | Presence |
|---|---|---|---|---|
| Total Score for Presence of Information Security (IS) Policy | ✓ | 0 | 7 | Very Low |
| | | 8 | 15 | Low |
| | | 16 | 30 | Medium |
| | | 31 | 45 | High |
| | | 46 | 60 | Very High |
| Total Score for Presence of Procedures and Controls | ✓ | 0 | 4 | Very Low |
| | | 5 | 8 | Low |
| | | 9 | 15 | Medium |
| | | 16 | 22 | High |
| | | 23 | 30 | Very High |
| Total Score for Presence of Administrative tools and Methods | ✓ | 0 | 2 | Very Low |
| | | 3 | 6 | Low |
| | | 7 | 12 | Medium |
| | | 13 | 18 | High |
| | | 19 | 25 | Very High |
| Total Score for Presence of Awareness Creation | ✓ | 0 | 5 | Very Low |
| | | 6 | 11 | Low |
| | | 12 | 22 | Medium |
| | | 23 | 33 | High |
| | | 34 | 45 | Very High |
| **TOTAL SCORE FOR PRESENCE OF ORGANIZATIONAL MEASURES** (Presence of IS Policy, Procedures, Administrative tools and Awareness creation) | ✓ | | | |

Table 5: Overall Information Systems Security Assessment Rating

| Presence of Technological Measures | Total Score for Presence of Organizational Measures | | Presence of Organizational Measures | Overall Assessment |
|---|---|---|---|---|
| Very High | 0 | 90 | Poor | Poor practices, organizational measures need immediate attention |
| | 91 | 130 | Needs Improvement | Good practice, but organizational measures need improvement |
| | 131 | 160 | Good | Very good practice |
| High | 0 | 80 | Poor | Poor practices, organizational measures need immediate attention |
| | 81 | 120 | Needs Improvement | Good practice, but organizational measures need improvement |
| | 121 | 160 | Good | Very good practice |
| Medium | 0 | 70 | Poor | Poor practices, technological measures need improvement and organizational measures need immediate attention |
| | 71 | 110 | Needs Improvement | Average practice, but organizational measures need improvement |
| | 111 | 160 | Good | Good practice, but technological measures need improvement |
| Low | 0 | 60 | Poor | Very poor practices, technological measures and organizational measures need urgent attention |
| | 61 | 100 | Needs Improvement | Poor practices, technological measures and organizational measures need immediate attention |
| | 101 | 160 | Good | Poor practices, technological measures need immediate attention |
| Very Low | 0 | 50 | Poor | Very poor practices, technological measures and organizational measures need urgent attention |
| | 51 | 90 | Needs Improvement | Poor practices, technological measures and organizational measures need immediate attention |
| | 91 | 160 | Good | Poor practices, technological measures need immediate attention |

Table 6: Implementation of Technological Measures by Types of Libraries

| Level of Implementation Technological Measures | | Type of Libraries | | |
|---|---|---|---|---|
| | | Special Library | Public Library | Total |
| Medium | Count | 2 | 0 | 2 |
| | % within Column | 5.0% | .0% | 4.0% |
| High | Count | **38** | **10** | **48** |
| | % within Column | **95.0%** | **100.0%** | **96.0%** |
| Total | Count | 40 | 10 | 50 |
| | % within Column | 100.0% | 100.0% | 100.0% |

The libraries have high presence of technological security controls to protect the hardware, workstations, servers, software, network and their physical facilities. This seems to affirm the views that the answers to security challenges have been technological (Volonino and Robinson 2004). However, the presence of data security measures in these libraries is average, and this might threaten the confidentiality, integrity and availability of data. The libraries therefore need to assure adequate controls are in place to prevent disclosure of confidential data (Table 7).

Table 7: Presence of Technological Measures in Malaysian Special and Public Libraries

| Presence of Technological Measures | | Very Low | Low | Medium | High | Very High | Total |
|---|---|---|---|---|---|---|---|
| Hardware Security | Count | 0 | 4 | 16 | 22 | 8 | 50 |
| | % | .0% | 8.0% | 32.0% | 44.0% | 16.0% | 100.0% |
| Software Security | Count | 0 | 0 | 22 | 28 | 0 | 50 |
| | % | .0% | .0% | 44.0% | 56.0% | .0% | 100.0% |
| Workstation Security | Count | 0 | 0 | 12 | 34 | 4 | 50 |
| | % | .0% | .0% | 24.0% | 68.0% | 8..0% | 100.0% |
| Network Security | Count | 0 | 0 | 14 | 35 | 1 | 50 |
| | % | .0% | .0% | 28.0% | 70.0% | 2.0% | 100.0% |
| Server Security | Count | 0 | 0 | 5 | 40 | 5 | 50 |
| | % | .0% | .0% | 10.0% | 80.0% | 10.0% | 100.0% |
| Data Security | Count | 0 | 0 | 34 | 15 | 1 | 50 |
| | % | .0% | .0% | 68.0% | 30.0% | 2.0% | 100.0% |
| Physical Security | Count | 0 | 0 | 15 | 35 | 0 | 50 |
| | % | .0% | .0% | 30.0% | 70.0% | 0.0% | 100.0% |

All the libraries in this study have high level of implementation for technological measures and were the majority have 5 years to 10 years experiences in ICT or computerization and received between 1% and 3% of budget for IS security. This imply that adequate financial resources for IS security and longer ICT adoption would most likely lead to the presence of good technological measures in their libraries ($x^2$ =50,000, df2, p = 0.001; $x^2$ =8.333, df2 , p = 0.016 respectively).

### (b) Levels of Implementation of Organizational Measures

The majority of the libraries (54.0%) in the pilot study have poor implementation of organizational security measures, and, 46.0% require improvement on the implementation of organizational measures (Table 8). This unsecured situation could result in a variety of security risks as most of today's security challenges are related to human and organizational aspects of security (Anderson 2007).

Table 8: Implementation of Organizational Measures by Types of Libraries

| Level of Implementation Organizational Measures | | Type of Libraries | | Total |
|---|---|---|---|---|
| | | Special Library | Public Library | |
| Poor | Count | 22 | 5 | **27** |
| | % within Column | 55.0% | 50.0% | **54.0%** |
| Needs Improvement | Count | 18 | 5 | **23** |
| | % within Column | 45.0% | 50.0% | **46.0%** |
| Total | Count | 40 | 10 | 50 |
| | % within Column | 100.0% | 100.0% | 100.0% |

The majority of libraries (72%) in this study demonstrate high presence of information security policies implementation. This finding corroborates the report made by Dimopoulos et al. (2004), which indicated that the practice of developing information security policies is becoming increasingly popular and may be considered a source of competitive advantage amongst security conscious practitioners.

However, the results show average emphasis on the presence of security procedures and controls, administrative tools and awareness creation activities (Table 9). Literature has highlighted the importance of creating supporting policies, standards, guidelines and procedures in an organization to ensure protection against human-created security problems such as detailing on the roles for security administrators and users to maintain the security of the systems (Dhillon 2001; Conklin et al. 2005).

Table 9: Presence of Organizational Measures in Malaysian Special and Public Libraries

| Presence of Organizational Measures | | Very Low | Low | Medium | High | Very High | Total |
|---|---|---|---|---|---|---|---|
| Information Security | Count | 0 | 0 | 8 | 36 | 6 | 50 |
| (IS) Policy | % | .0% | .0% | 16.0% | 72.0% | 12.0% | 100.0% |
| Procedures and | Count | 0 | 0 | 25 | 25 | 0 | 50 |
| Controls | % | .0% | .0% | 50.0% | 50.0% | .0% | 100.0% |
| Administrative tools | Count | 0 | 6 | 37 | 7 | 0 | 50 |
| and Methods | % | .0% | 12.0% | 74.0% | 14.0% | .0% | 100.0% |
| Awareness Creation | Count | 0 | 0 | 37 | 13 | 0 | 50 |
| | % | .0% | .0% | 74.0% | 26.0% | .0% | 100.0% |

This study found that libraries with less than 10 years experiences of ICT adoption (46.4%) reported lower implementation of organizational measures ($x^2$ =9.819, df 2, p=0.007).. Similarly, a larger proportion of libraries without wireless connection (87.5%) were among the libraries with poor level of organizational measures. Also, libraries (73%) which received less than 1% of financial assistance for ISS were among the libraries with poor level of organizational measures ($x^2$ =16.318, df 2, p=0.000). When comparing the status of organizational and technological measures, it is apparent that a large numbers of libraries in the sample which have high implementation of technological measures do not necessarily have good practice in organizational measures.

**(c) Overall Status of Library ISS Measures**

Overall, the results indicate that 46.0% of the libraries in this survey have good practice of ISS measures but their organizational practices require improvement (Table 10). Whereas 50.0% of libraries have poor practices of ISS measures and need to pay immediate attention to their organizational measures. Only a small number of libraries (4.0%) have poor practices where their technological measures need improvement and organizational measures require immediate attention. This present findings is consistent with other research which found that a significant portion of human-created security problems resulted from poor organizational security practices, such as users not following established security policies or processes, and a lack of security policies, procedures, or training within the organization (Conklin et al. 2005). As people are the weakest link in any security-related process, libraries need to addresses issues on user education, awareness, and training on information security policies and procedures that affect them (Merkow and Breithaupt 2005).

Table 10: Overall Status of ISS Practices in Malaysian Academic Libraries

| Overall Status of Information Systems Security Practices | | Type of Libraries | | |
| --- | --- | --- | --- | --- |
| | | Special Library | Public Library | Total |
| Good practice but organizational measures need improvement | Count | 18 | 5 | 23 |
| | % within Column | 45.0% | 50.0% | 46.0% |
| Poor practices, organizational measures need immediate attention | Count | 20 | 5 | 25 |
| | % within Column | 50.0% | 50.0% | 50.0% |
| Poor practices, technological measures need improvement and organizational measures need immediate attention | Count | 2 | 0 | 2 |
| | % within Column | 5.0% | .0% | 4.0% |
| Total | Count | 40 | 10 | 50 |
| | % within Column | 100.0% | 100.0% | 100.0% |

**CONCLUSION**

It was evident that the level of implementation of technological measures in Malaysian public and special libraries is high. This may be related to the provision of sufficient financial support and years of experiences in ICT or computerization. However, the implementation level of organizational measures in these libraries is considered poor. This may be due to the over-emphasis on technology as the sole solution to all security problems and needs to be investigated further. The libraries need to factor in organizational issues to improve their IS security practices.

New technologies are forcing libraries to be particularly attentive to changes of IS risks associated and the controls used to secure the IS. Thus, maintaining the library's ISS means not only ensuring that library patrons are provided access to information but also maintaining and monitoring the library's hardware, software, and security, guided by documented policies and procedures (Breeding 2003). Libraries should also maintain the privacy of their information assets such as the library's financial information, patrons' circulation information, and passwords to access the library systems. Patrons should be given appropriate access to the library computers, web sites, databases, and servers based on the principle of least privilege, which refers to only the privileges that they need to perform their specific jobs or tasks in order to protect data integrity from any breaches. Libraries should also implement good backup policies and recovery procedures to ensure their data and services via information systems can be accessed and shared in a convenient way whenever it is needed and data can be restored quickly during downtime. This study is limited by the small sample size of the data collected. In future other contributing factors need to be investigated such as the impact of local environment, ethics and culture that may be related to the degree of implementation ISS in a variety of library settings.

**REFERENCES**

Al-Salihy, W., Ann, J. and Sures, R. 2003. Effectiveness of information systems security in IT organizations in Malaysia. *Proceedings of 9th Asia-Pacific Conference on Communication.* 21-24 Sept 2003.Vol.2:716-720.

Anderson, K. 2007. Convergence: A holistic approach to risk management. *Network Security*, Vol.5: 4-7.

Berghel, H. 2005. The Two Sides of RoI: Return on Investment vs. Risk of Incarceration, *Communications of the ACM,* Vol.48, no.4: 15-20.

Breeding, M. 2003. Protecting your library's data. Computers in Libraries. Available at http://www.librarytechnology.org/diglibfulldisplay.pl?SID=20110116654235839&ode=bib&RC=10343&Row=31&.

Conklin, W.A., White, G.B., Cothren, C., William, D. and Davis, R.L. 2005. *Principles of Computer Security: Security+ and Beyond*. Illinois: McGrawHill Technology Education.

Dimopoulos, V., Furnell, S., Barlow, I. and Lines, B. 2004. Factors affecting the adoption of IT risk analysis. In *Proceedings of 3rd European Conference on Information Warfare and Security,* Royal Holloway, University of London, UK, 28-29 June 2004.

Doherty, N.F. and Fulford, H. 2006. Aligning the information security policy with the strategic information systems plan. *Computers & Security*, Vol.25, no.1: 55-63.

Dhillon, G. 2001. Challenges in Managing Information Security in the New Millennium, In, Dhillon, G. (Ed.), *Information Security Management: Global Challenges in the New Millennium*, Hershey, PA: Idea Group Publishing: pp. 1-8.

EDUCAUSE/Internet2 Security Task. 2004. *The Information Security Governance (ISG) Assessment Tool for Higher Education*. Available at http://net.educause.edu/ir/library/pdf/SEC0421.pdf.

Eisenberg, J. and Lawthers, C. 2008. Library Computer and Network Security: Library Security Principles. Infopeople Project. Available at: http://www.infopeople.org/resources/ security/basics/index.html.

Guel, M.D. 2007. *A Short Primer for Developing Security Policies*. Available at: http://www.sans.org/resources/ policies/Policy_Primer.pdf.

Gupta, M. and Sharman, R. 2008. *Social and Human Elements of Information Security: Emerging Trends and Countermeasures*. Hershey: PA, IGI Global.

Guttman, B. and Roback, E. 1995. *An Introduction to Computer Security: The NIST Handbook. U.S. National Institute of Standards and Technology, NIST Special Publication 800-1.* Available at http://csrc.nist.gov/publications/nistpubs/800-12/handbook.pdf.

Hagen, J.M., Albrechtsen, E. and Hovden, J. 2008. Implementation and effectiveness of organizational information security measures. *Information Management & Computer Security,* Vol.16, no.4: 377-397.

Hone, K. and Eloff, J.H.P. 2002. Information security policy – what do international security standards say? *Computers & Security,* Vol.21, no.5: 402-409.

Information-technology Promotion Agency. 2008. *Information Security Management Benchmark (ISM-Benchmark).* Available at http://www.ipa.go.jp/security/ english/benchmark/.../Howtouse_ISM_Benchmark. pdf.

INTOSAI. 1995. *Information System Security Review Methodology: A Guide for Reviewing Information System Security in Government Organizations.* Available at http://www.issai.org/media(421,1033)/ISSAI_5310_E.pdf.

MacCallum, R.C., Widaman, K.F., Zhang, S. and Hong, S. 1999. Sample size in factor analysis. *Psychological Methods:* 84-99.

MacCallum, R.C., Widaman, K.F., Preacher, K.J. and Hong, S. 2001. Sample size in factor analysis: The role of model error. *Multivariate Behavioral Research*, Vol.36:611-637.

Malhotra, N.K. 2004. *Marketing research: an applied orientation*, 4[th] ed, Pearson Prentice Hall. Available at http://www.sunyit.edu/~barnesj2/mkt550/malhorta19.ppt.

Merkow, M. and Breithaupt, J. 2005. *Principles of Information Security: Principles and Practices.* Pearson Prentice Hall: Upper Saddle River, New Jersey.

Newby, G.B. 2002. *Information Security for Libraries.* Available at http://www. petascale.org/papers/library-security.pdf.

Pipkin, D.L. 2000. *Information Security: Protecting the Global Enterprise.* Upper Saddle River, NJ: Prentice Hall.

Powell, A. and Gillet, M. 2007. Controlling Access in the Electronic Library, *Ariadne*, Vol.7. Available at http://www.ariadne.ac.uk/issue7/access- control.

Samy, G.N., Rabiah, A. and Zuraini, I. 2009. Security threats in healthcare information systems: A preliminary study. In: *Fifth International Conference on Information Assurance and Security*. IEEE Computer Society, 18-20 August, 2009, Xian, China.

Scarfone, K., Souppaya, M., Cody, A. and Orebaugh, A. 2008. *Technical Guide to Information Security Testing and Assessment*. Technical Report Spec. Publ. 800-11, (U.S. Department of Commerce, National Institute of Standards and Technology). Available at http://csrc.nist.gov/publications/nistpubs/800-115/SP800-115.pdf.

Siponen, M.T. and Oinas-Kukkonen, H. 2007. A review of information security issues and respective research contributions. *The Database for Advances in Information Systems*, Vol.38, no.1: 60-81.

Suhazimah, D. 2007. *The antecedents of information security maturity in Malaysian public service organizations*. Ph.D. thesis. Faculty of Business and Administration, University of Malaya, Malaysia.

Sundt, C. 2006. Information security and the law, *Information Security Technical Report*, Vol.11, no.1: 2-9.

Thiagarajan, V. 2002. *Information Security Management BS 7799.2:2002 Audit Check List for SANS.*

Volonino, L. and Robinson, S. R. 2004. *Principles and Practice of Information Security: Protecting Computers from Hackers and Lawyers.* Pearson Education: Upper Saddle River, p.1

Von Solms, B. 2000. Information security – the third wave? *Computers & Security*, Vol.19, no.7: 615-620.

Westby, J.R. and Allen, J.H. 20070. *Governing for Enterprise Security (GES) Implementation Guide* (CMU/SEI-2007-TN-020), Pittsburgh, PA., Software Engineering Institute, Carnegie Mellon University. Available at http://www.cert.org/archive/pdf/ 07tn020.pdf.

Weise, J. and Martin, C.R. 2001. *Sample Data Security Policy and Guidelines Template,* Sun Blue Prints, 2001) OnLine. Available at http://www.sun.com/blueprints/tools/ samp_sec_pol.pdf.

Williams, R.L. 2001. *Computer and network security in small libraries: A guide for planning*. Texas State Library & Archives Commission). Available at http://www.tsl.state.tx.us/ld/pubs/compsecurity.

Yeh, Q. and Chang, A.J. 2007. Threats and countermeasures for information system security: A cross-industry study. *Information & Management*, Vol.44: 480-491.

Zhao, N. 2009. *The Minimum Sample Size in Factor Analysis*. Available at http://www.encorewiki.org/display/~nzhao/The+Minimum+Sample+Size+in+Factor+Analysis.

## Appendix 1: The Library Information System Security Assessment Components

**The following is a list of Information Systems (IS) safeguarding measures.**
Please tick **(√)** in the box to indicate the level of implementation in your library based on index below:
**1** - Not implemented    **2** - Only some part has been implemented    **3** - Implemented but has not been reviewed
**4** - Implemented and reviewed on regular basis    **5** - Fully implemented and recognised as good example for other libraries

| a) | **Hardware security** | | | | | |
|---|---|---|---|---|---|---|
| 1.1 | CCTV, visual camera, magnetic detection system and electronic anti-theft system at strategic places, public computer areas and server areas. | ① | ② | ③ | ④ | ⑤ |
| 1.2 | Emergency power sources and alternative communication lines (telephone lines and generators) | ① | ② | ③ | ④ | ⑤ |
| 1.3 | Locks, security cables, locked cable trays, metal cages or anchoring devices to improve the security of hardware equipments. | ① | ② | ③ | ④ | ⑤ |
| 1.4 | Periodical remote mirroring or file mirroring to back up disk drives. | ① | ② | ③ | ④ | ⑤ |
| | **Total Score (a)** | | | | | |
| b) | **Software Security** | | | | | |
| 1.5 | Anti-spyware software to detect and remove any spyware threats. | ① | ② | ③ | ④ | ⑤ |
| 1.6 | Anti-phishing solutions to prevent phishing attacks. | ① | ② | ③ | ④ | ⑤ |
| 1.7 | Cleanup software to erase files or settings left behind by a user. | ① | ② | ③ | ④ | ⑤ |
| 1.8 | Desktop security software at application and operating level to monitor, restrict usage or disable certain features of the workstations. | ① | ② | ③ | ④ | ⑤ |
| 1.9 | Automate the process of installing an application or updates to workstations on a network. | ① | ② | ③ | ④ | ⑤ |
| 1.10 | ID management software to automate administrative tasks such as resetting user passwords and enabling users to reset their own passwords. | ① | ② | ③ | ④ | ⑤ |
| 1.11 | Menu replacement software to replace the standard windows desktop interfaces and provide control on timeouts, logging and browsing activities. | ① | ② | ③ | ④ | ⑤ |
| 1.12 | Multi-user operating systems and application software to allow concurrent access by multiple users of a computer. | ① | ② | ③ | ④ | ⑤ |
| 1.13 | Automatic debugging and tests to remove any defects from new software or hardware components. | ① | ② | ③ | ④ | ⑤ |
| 1.14 | Rollback software to keep track and record any changes made to the computers and allow the system to be restored to its original state from any chosen point in time. | ① | ② | ③ | ④ | ⑤ |
| 1.15 | Single sign-on system for a user authentication and authorization to access all computers and systems without the need to enter multiple passwords. | ① | ② | ③ | ④ | ⑤ |
| 1.16 | Spam filtering software to detect unwanted spam emails from getting into users' inboxes. | ① | ② | ③ | ④ | ⑤ |
| 1.17 | Systems recovery to rebuild, repair the library computer systems after disaster or crash. | ① | ② | ③ | ④ | ⑤ |
| 1.18 | Timer software to control the amount of time a patron can use a workstation. | ① | ② | ③ | ④ | ⑤ |
| 1.19 | User entrance log to record and monitor user logs which are regularly analyzed. | ① | ② | ③ | ④ | ⑤ |
| 1.20 | Web filtering software to prevent access to inappropriate materials or sites. | ① | ② | ③ | ④ | ⑤ |
| | **Total Score (b)** | | | | | |
| c) | **Workstation Security** | | | | | |
| 1.21 | All office productivity software and browsers for the workstations/laptops are configured to receive updates in a timely manner. | ① | ② | ③ | ④ | ⑤ |
| 1.22 | An application firewall is used for mobile laptops that connect to the library's LAN. | ① | ② | ③ | ④ | ⑤ |
| 1.23 | The computer's BIOS are secured in order to create a secure public access computer. | ① | ② | ③ | ④ | ⑤ |
| 1.24 | User identification and authentication are required before logging into the library's workstations, laptops screensavers, library network or campus network. | ① | ② | ③ | ④ | ⑤ |
| 1.25 | Virus protection programs, configuration settings and security software programs are installed for web browsers and email programs. | ① | ② | ③ | ④ | ⑤ |
| | **Total Score (c)** | | | | | |
| d) | **Network Security** | | | | | |
| 1.26 | Antivirus software and desktop security software to receive regular updates to protect the internal network from any security breaches. | ① | ② | ③ | ④ | ⑤ |
| 1.27 | Digital signatures are used to assure the authenticity of any electronic documents sent via the library's network (use of passwords, private and public key encryption or digital certificates) | ① | ② | ③ | ④ | ⑤ |
| 1.28 | Firewall to protect the internal network from external threats. | ① | ② | ③ | ④ | ⑤ |
| 1.29 | Firewall with virtual private network is installed for remote and wireless access connections. | ① | ② | ③ | ④ | ⑤ |
| 1.30 | Limitation of connection time is performed via configuration routines to control and restrict access for the library's high-risk applications or databases. | ① | ② | ③ | ④ | ⑤ |
| 1.31 | Public and staff's local area networks (LANs) are physically separated by means of separate cabling for each network to provide alternative circuit. | ① | ② | ③ | ④ | ⑤ |
| 1.32 | Server segregation/perimeter network (DMZ) by using firewalls and some other network access control devices to separate systems that are at a relatively high risk from unsecured network. | ① | ② | ③ | ④ | ⑤ |
| 1.33 | The network is segmented with a router to increases the bandwidth available to each user and reduce the congestions or collisions of the library's network. | ① | ② | ③ | ④ | ⑤ |
| 1.34 | Wireless security products to secure the library wireless network. (use of default passwords on wireless access points, network ID, wireless intrusion detection systems, wired equivalency protocol (WEP) encryption, MAC address filtering or virtual private networking (VPN)) | ① | ② | ③ | ④ | ⑤ |
| | **Total Score (d)** | | | | | |

| **e)** | **Server Security** | | | | | |
|---|---|---|---|---|---|---|
| 1.35 | Anti-virus software on servers and anti-virus definition files are kept up-to-date. | ① | ② | ③ | ④ | ⑤ |
| 1.36 | Authentication systems to prevent unauthorized access to the library's server. | ① | ② | ③ | ④ | ⑤ |
| 1.37 | Implement fault tolerance to ensure if one system fails, then there is a backup system that immediately takes over. | ① | ② | ③ | ④ | ⑤ |
| 1.38 | Firewalls to protect the library network from unwarranted intrusion. | ① | ② | ③ | ④ | ⑤ |
| 1.39 | Intrusion detection software and host auditing software are installed to monitors the servers or computers for signs of intrusion. | ① | ② | ③ | ④ | ⑤ |
| 1.40 | Regular backups for the data, hard copy of server hardware specifications, installation information, installation software and passwords are regularly performed and stored at an offsite location. | ① | ② | ③ | ④ | ⑤ |
| 1.41 | Server logs are reviewed periodically by using a log file monitor utility to monitor any signs of intrusion or security violations. | ① | ② | ③ | ④ | ⑤ |
| 1.42 | Restrict access to the file system in a server to the directory structure using file or directory permissions. | ① | ② | ③ | ④ | ⑤ |
| 1.43 | Library servers' operating systems and applications are hardened to protect from any vulnerabilities. | ① | ② | ③ | ④ | ⑤ |
| 1.44 | The server is placed in a secure location, such as in a lockable cage, a locked room and place it with environmental controls. | ① | ② | ③ | ④ | ⑤ |
| | **Total Score (e)** | | | | | |
| **f)** | **Data Security** | | | | | |
| 1.45 | Attributes for each removable media applications in your library are properly recorded and the media are kept from any unauthorized devices from accessing, running or transferring data to your library workstations and network. (USB thumb drives, tapes, CDs, DVDs, disks, drives, etc). | ① | ② | ③ | ④ | ⑤ |
| 1.46 | Combination of authentication systems to restrict access to library data and resources based on a variety of access rights (user identification, passwords or biometrics system) | ① | ② | ③ | ④ | ⑤ |
| 1.47 | Disposable of unused media and sensitive media are properly managed to maintain an audit trail. | ① | ② | ③ | ④ | ⑤ |
| 1.48 | Enforced path is created between a user terminal and other library services to reduce the risk of unauthorized access. | ① | ② | ③ | ④ | ⑤ |
| 1.49 | Event logging or log management software to ensure the library computer security records are stored in sufficient detail for an appropriate period of time. (records for security incidents, policy violations, fraudulent activities and operational problems) | ① | ② | ③ | ④ | ⑤ |
| 1.50 | Fraud detection and prevention measures to control fraudulent activity and disclosure of information (use of address verification system, proprietary encryption, internal intrusion detection system, multiple login monitoring, password verification on transactions or data access controls) | ① | ② | ③ | ④ | ⑤ |
| 1.51 | Public key infrastructure (PKI) to secure the exchange of personal data via the library network and Internet. (use of public and private cryptography key pair). | ① | ② | ③ | ④ | ⑤ |
| 1.52 | RFID tags to manage and secure the library collection as well as to track attendance and prevent unauthorized access into the library building. | ① | ② | ③ | ④ | ⑤ |
| 1.53 | Systematic approaches conducted in-house or outsourced to a service provider to address the library's vulnerabilities (vulnerability discovery, prioritization, remediation, dynamic protection, verification and customized reporting). | ① | ② | ③ | ④ | ⑤ |
| 1.54 | Use of cryptography techniques, hardware & software tokens and single sign on systems to control data access to the library internal and remote computer systems. | ① | ② | ③ | ④ | ⑤ |
| 1.55 | Use of password protection of user accounts, anti-virus software, firewalls, wireless network protections, intrusion detection systems and Internet Protocol Virtual Private Networks/IP VPNs to ensure data insert and sent from one end of a transaction arrives unaltered at the other end. | ① | ② | ③ | ④ | ⑤ |
| 1.56 | Vital library's business information or records are regularly backed up. ( inventory records, patrons' data, library databases, production servers and critical network components and backup media). | ① | ② | ③ | ④ | ⑤ |
| 1.57 | Web access management systems to manage and validate user access to devices, applications and library resources (authentication management, single sign-on convenience, audit or reporting systems). | ① | ② | ③ | ④ | ⑤ |
| | **Total Score (f)** | | | | | |
| **g)** | **Physical and Environmental Security** | | | | | |
| 1.60 | Air conditionings to stabilize the temperature & humidity within the library building. | ① | ② | ③ | ④ | ⑤ |
| 1.61 | Earthquake early warning system to alert library staff and patrons prior to damaging ground shaking. | ① | ② | ③ | ④ | ⑤ |
| 1.62 | Flood detector to provide an early warning of developing floods in a library. | ① | ② | ③ | ④ | ⑤ |
| 1.63 | Lightning protectors and surge protectors to protect any valuable machines or equipments from lightning strikes, voltage spikes and surges. | ① | ② | ③ | ④ | ⑤ |
| 1.64 | Security guards to monitor people entering and leaving the library buildings and sites. | ① | ② | ③ | ④ | ⑤ |
| 1.65 | Use of automatic sprinkler systems, smoke detectors, fire extinguishers and fireproof installations in the library buildings and areas adjacent to library's key assets to detect and prevent fires, toxic chemical spills and explosions. | ① | ② | ③ | ④ | ⑤ |
| 1.66 | Use of magnetic stripe swipe cards, electronic lock, proximity cards, bar code card or biometrics to secure and control access to restricted library areas. | ① | ② | ③ | ④ | ⑤ |
| 1.67 | Warning signs, fencing, vehicle height-restrictors, site lightings and trenches around the library areas to provide initial layer of security for a library building. | ① | ② | ③ | ④ | ⑤ |
| 1.68 | Wireless gates, biometrics or other user identifications and authentication forms at the library main entrances, exits and public access areas in the library building. | ① | ② | ③ | ④ | ⑤ |
| | **TOTAL SCORE FOR** *(a+ b+ c+ d+ e+ f+ g)* | | | | | |
| **2.0** | **Information Security Policy** | | | | | |
| 2.1 | Back-ups and off-site storage policies for your library data, media or materials that contain sensitive information. | ① | ② | ③ | ④ | ⑤ |
| 2.2 | Data classification, retention and destruction policies for your library data, media or materials that contain sensitive information. | ① | ② | ③ | ④ | ⑤ |
| 2.3 | Identity management policies for library Information Systems user registration and password management. | ① | ② | ③ | ④ | ⑤ |
| 2.4 | Job responsibility policy for individual employee responsibilities related to the library IS security practices. | ① | ② | ③ | ④ | ⑤ |
| 2.5 | Policies on access control, authentication and authorisation practices for using the library Information Systems. | ① | ② | ③ | ④ | ⑤ |
| 2.6 | Policies on protection of library IS assets to protect your library's hardware, software, data and people. | ① | ② | ③ | ④ | ⑤ |
| 2.7 | Secure disposal policies for library data, media or materials that contain sensitive information. | ① | ② | ③ | ④ | ⑤ |
| 2.8 | Polices on reporting, notification and response of Information Systems security events to affected parties such as individuals, law enforcement, campus or parent organisations. | ① | ② | ③ | ④ | ⑤ |
| 2.9 | Policies on acceptable use of wireless devices in your library such as laptops and hand phones. | ① | ② | ③ | ④ | ⑤ |
| 2.10 | Policies on acceptable use of workstations, e-mails, databases, intranet and Internet in your library. | ① | ② | ③ | ④ | ⑤ |
| 2.11 | Policies on managing privacy and confidentiality issues, including breaches of personal information. | ① | ② | ③ | ④ | ⑤ |
| 2.12 | Policies on sharing, storing and transmitting of library data via ISPs, external networks or contractors' systems. | ① | ② | ③ | ④ | ⑤ |
| | **TOTAL** | | | | | |
| **3.0** | **Information Security Policy** | | | | | |
| 3.1 | Controls and disciplinary procedures if a library staff or patrons breach the IS security policies or rules. (verbal warning, written warning, suspension and dismissal). | ① | ② | ③ | ④ | ⑤ |
| 3.2 | Procedures for handling library sensitive data and personal data of library patrons to prevent errors, unauthorised disclosure or misuse by those who handle it. | ① | ② | ③ | ④ | ⑤ |
| 3.3 | Procedures for non-disclose agreement or confidentiality agreement to all library staff and patrons to protect any type of confidential and proprietary information. | ① | ② | ③ | ④ | ⑤ |
| 3.4 | Procedures for update and review existing information security policies. | ① | ② | ③ | ④ | ⑤ |
| 3.5 | Procedures on the intellectual property rights and copyrights in controlling and protecting any digital works or resources that are stored, transmitted, accessed, copied or downloaded via the library IS. | ① | ② | ③ | ④ | ⑤ |
| 3.6 | Procedures which list all requirements with regard to outsourcing any library Information Systems service or activities. | ① | ② | ③ | ④ | ⑤ |
| | **TOTAL** | | | | | |
| **4.0** | **Information Security Policy** | | | | | |
| 4.1 | Procedure for owner accountability to ensure appropriate protection is maintained for each library IS asset. (e.g. information assets, software assets, physical assets and library services). | ① | ② | ③ | ④ | ⑤ |
| 4.2 | Procedures for the development and implementation of risk analysis to protect your library from all types of threats. (Performance of assets analysis, threat analysis, annual loss expectancy analysis, identification and evaluation of security measures). | ① | ② | ③ | ④ | ⑤ |
| 4.3 | Procedures on handling, reporting, notification and response of IS security events to affected parties such as individuals, law enforcement, campus or parent organisation. | ① | ② | ③ | ④ | ⑤ |
| 4.4 | Procedures related to asset classification in order to organise it according to its importance and sensitivity to loss. (unclassified, confidential, secret and top secret) | ① | ② | ③ | ④ | ⑤ |
| 4.5 | Regular internal and external audits programs appropriate for your library's Information Systems size, complexity of activities, scope of operations, risk profile and compliance with the relevant standards. | ① | ② | ③ | ④ | ⑤ |
| | **TOTAL** | | | | | |
| **5.0** | **Awareness Creation** | | | | | |
| 5.1 | All staff and patrons at various levels are made aware of their responsibilities with regard to protecting the library's Information Systems' security and trained to report any security breach incidences. | ① | ② | ③ | ④ | ⑤ |
| 5.2 | All staff and patrons at various levels receive appropriate information security trainings and education. | ① | ② | ③ | ④ | ⑤ |
| 5.3 | All staff and patrons at various levels receive regular updates on your library Information Systems' policies and procedures. | ① | ② | ③ | ④ | ⑤ |
| 5.4 | Information security awareness trainings have become mandatory to all staff and patrons at various levels. | ① | ② | ③ | ④ | ⑤ |
| 5.5 | Risk assessment approach exists and follows a defined process that is documented. | ① | ② | ③ | ④ | ⑤ |
| 5.6 | Staff and patrons at various levels are trained to monitor and handle the library's Information Systems on their own. | ① | ② | ③ | ④ | ⑤ |
| 5.7 | There are balanced set of key performance indicators (KPIs) and metrics used to provide the real insight into the effectiveness of security awareness programs. | ① | ② | ③ | ④ | ⑤ |
| 5.8 | There are positive supports and commitments from the top management to coordinate the implementation of Information Systems' security controls in your library. (e.g. via allocation of budget, strong interest and active involvements). | ① | ② | ③ | ④ | ⑤ |
| 5.9 | Threats that could harm and adversely affect critical operations of your library Information Systems' security are identified and up dated regularly. | ① | ② | ③ | ④ | ⑤ |
| | **TOTAL** | | | | | |